\documentclass[prb,twocolumn,showpacs,superscriptaddress]{revtex4}% Physical Review B

\usepackage{graphicx}
\usepackage{dcolumn}

\usepackage{eucal}
\usepackage[dvips]{epsfig}
\usepackage{changebar}
\usepackage{amssymb}
\usepackage{amsfonts}

\usepackage{amsmath}
\newcommand{\EW}[1]{\left\langle{#1}\right\rangle}

\newcommand{\GREEN}[3]{\left\langle\hspace{-.4ex}\left\langle#1;
      #2\right\rangle\hspace{-.4ex}\right\rangle{#3}}

\begin{document}

%%%%%%%%%%%%%%%%%%%%%%%%%%%%%%%%%%
\newcommand{\csp}{\;,\qquad\qquad} %???
%%%%%%%%%%%%%%%%%%%%%%%%%%%%%%%%%%%%%%%%%%%%%%%%%%%%%%%%%%%%%%%%%%%%%%
\title{Carrier induced ferromagnetism in diluted local-moment systems}

\author{Guixin Tang}
% \email{tangguixin@googlemail.com}
\affiliation{Institut f\"ur Physik, Humboldt-Universit\"at zu Berlin,
 Newtonstra{\ss}e 15, D-12489 Berlin, Germany}
\affiliation{Department of Physics, Harbin Institute of
Technology, D-150001 Harbin, People's Republic of China}
\author{Wolfgang Nolting}
% \email{wolfgang.nolting@physik.hu-berlin.de}
\affiliation{Institut f\"ur Physik, Humboldt-Universit\"at zu Berlin,
 Newtonstra{\ss}e 15, D-12489 Berlin, Germany}

\date{\today}

\begin{abstract}

The electronic and magnetic properties of concentrated and diluted ferromagnetic semiconductors are investigated by using the Kondo lattice model, 
which describes an interband exchange coupling between itinerant conduction electrons and localized magnetic moments. 
In our calculations, the electronic problem and the local magnetic problem are solved separately. 
For the electronic part an interpolating self-energy approach together with a coherent potential approximation (CPA) treatment 
of a dynamical alloy analogy is used to calculate temperature-dependent quasiparticle densities of states 
and the electronic self-energy of the diluted local-moment system. 
For constructing the magnetic phase diagram we use a modified RKKY theory by mapping the interband exchange to an effective Heisenberg model. 
The exchange integrals appear as functionals of the diluted electronic self-energy being therefore temperature- and 
carrier-concentration-dependent and covering RKKY as well as double exchange behavior. 
The disorder of the localized moments in the effective Heisenberg model is solved by a generalized locator CPA approach. The main results are: 
1) extremely low carrier concentrations are sufficient to induce ferromagnetism; 
2) the Curie temperature exhibits a strikingly non-monotonic behavior as a function of carrier concentration with a distinct maximum; 
3) $T_C$ curves break down at critical $n/x$ due to antiferromagnetic correlations and 
4) the dilution always lowers $T_C$ but broadens the ferromagnetic region with respect to carrier concentration. 

\end{abstract}


\maketitle

%%%%%%%%%%%%%%%%%%%%%%%%%%%%%%%%%%%%%%%%%%%%%%%%%%%%%%%%%%%%%%%%%%%%%%
\section{Introduction}
\label{sec:Introduction}

Diluted Magnetic Semiconductors (DMS) based on III-V and II-VI semiconductors doped with magnetic impurities such as Mn 
have attracted great experimental and theoretical interest in recent years due to their potential in spintronics applications\cite{Jungwirth06}. 
One of the most widely studied DMS is Ga$_{1-x}$Mn$_x$As of which Curie temperature as high as $T_C = 173 K$ has been observed\cite{Jungwirth06}. 
These materials indeed exhibit a very striking correlation between the transport and magnetic properties. 
Ideally, each Mn dopant atom represents an acceptor that introduces a local spin and a hole carrier. 
The ferromagnetism is driven by a charge-carrier mediated mechanism as a consequence of 
an interband exchange interaction between the localized magnetic moments and the carrier (hole) spins. 
Empirically, one finds that the hole density in DMSs is significantly smaller than the doping concentration due to carrier compensation etc. 
In addition, the measured saturated magnetization is observed to be much less than the nominal concentration of Mn ions\cite{Potashnik02}. 
Experimentally, the remanence magnetization and the Curie temperature $T_C$ are both enhanced by post-growth annealing of the samples, 
which changes positions of defects and the hole concentration\cite{Fiete0501,Fiete0502}. 
So the magnetism in DMSs is heavily affected by material issues, e.g. very disordered impurity-lattice-location, carrier compensation 
and the coupling mechanism between the localized Mn spins and the itinerant holes, as well as a complicated band structure. 
Therefore, it is impossible to study these materials without making some approximations. 

There are various theoretical approaches attempting to understand DMS physics. 
One class of models neglects interband coupling effects and only investigates the disorder of 
the localized moments\cite{Zhou04,Hilbert04,Tang0601,Tang0602}. 
Some models introduce the interband coupling between the itinerant carriers and the localized moments and study 
the effect of the charge carriers on the magnetic phase diagram
\cite{Nolting04,Fiete03,Alvarez03,Alvarez02,Mayr02,Bouzerar03,Bouzerar02,Chudnovskiy02,Berciu01}. 
If the ferromagnetic ordering is mediated by `free' carriers, its origin is assumed to be of RKKY type\cite{Zarand02}. 
However, further results show that RKKY interaction between the localized spins 
is insufficient\cite{Fiete0501,Fiete0502,Li05,Inoue03,Kudrnovsky04,Bouzerar06}. 
Korringa-Kohn-Rostock CPA (KKR-CPA) calculations\cite{Sato04,HAkai01} 
show that the ground state of Ga$_{1-x}$Mn$_x$As is half-metallic. 
The $\mathbf{k \cdot p}$ approach\cite{Jungwirth06B,Brey03,Yang03,Schliemann03} uses the $p-d$ exchange interaction between Mn and hole spins, 
in which the states near the Fermi energy are mainly assumed to have the character of the host semiconductor valence band. 
Moreover, dynamical mean-field theory (DMFT) \cite{SarmaDMFT,Aryanpour05,Majidi06} is used to investigate 
the carrier spin polarization of the impurity band and the Curie temperature of Ga$_{1-x}$Mn$_x$As. 
In addition to these theories, ab-initio approaches\cite{JLXu05,Silva05,XYCui05,Raebiger05}, 
spin-wave theories\cite{Konig00,Berciu02,Singh03,Hilbert05}, 
Monte Carlo (MC) simulations\cite{Mayr02,Bergqvist0405,Kennett02,Calderon02,Schliemann01} 
have been applied to understand the physics of DMSs. 

However, both interband-coupling effects and the disorder effects on the ferromagnetic phase transition 
have not been completely understood so far. 
Instead, the different theoretical models have tended to concentrate on different aspects of the problem. 
In this paper, we will use the Kondo lattice model (KLM) to investigate effects of the interband coupling 
and the disorder on the magnetic and electronic properties of diluted local-moment systems such as DMS. 
The positional disorder of the magnetic ions and the interband exchange coupling of charge carriers 
to the disordered local moments are treated on the same level and in the same theoretical framework. 
Here, we neglect the Coulomb disorder potential due to the substitutional Mn impurities. 
As a model study, the concentration of magnetic ions is not limited to $x \sim 0.05$ as in real DMS. 
We investigate instead the whole concentration range $x=0 \cdots 1$. 
The paper is expected to provide the \textit{qualitative} trends of the interconnected effects 
of interband coupling and disorder on the magnetic properties of diluted ferromagnetic semiconductors. 
Quantitative agreement with experiment will require treating both low magnetic ion concentration 
and defect correlations in a realistic way\cite{Potashnik02}. 

The article is organized as follows. 
The theoretical methods are described in section \ref{sec:Model}. 
Section \ref{sec:Numerical_Studies} is concerned with the numerical results and a discussion of them. 
Section \ref{sec:Summary} concludes the article with a summary. 

%%%%%%%%%%%%%%%%%%%%%%%%%%%%%%%%%%%%%%%%%%%%%%%%%%%%%%%%%%%%%%%%%%%%%%%%%%%%%%%%
\section{MODEL}\label{sec:Model}

The Kondo lattice model (KLM) is surely one of the most intensively discussed models over the past years in the field of magnetism. 
It describes the interaction between itinerant carriers (spin $\boldsymbol{\sigma}_i$) 
in a partially filled energy band and localized magnetic moments (spin $\mathbf{S}_i$) at certain lattice sites: 
\begin{equation}\label{eq:Hsf}
  H_{sf}=-J\sum_{i}\boldsymbol{\sigma}_i\cdot\mathbf{S}_i \;,
\end{equation}
where the index $i$ refers to the lattice site $\mathbf{R}_i$ and $J$ is the coupling between the itinerant carriers and the localized moments. 
The respective model Hamiltonian consists of two parts: 
\begin{equation}\label{eq:Hamil}
  H=H_s+H_{sf} \;.
\end{equation}
$H_s$ describes itinerant carriers in an energy band of finite width 
\begin{equation}\label{eq:Hs}
  H_s=\sum_{ij\sigma}T_{ij}c^{\dagger}_{i\sigma}c^{}_{j\sigma} \;,
\end{equation}
where $T_{ij}$ is the hopping integral, $c^{\dagger}_{i\sigma}(c_{i\sigma})$ creates (annihilates) a carrier on the site $i$ 
and $\sigma=\uparrow,\downarrow$ is the spin projection. 
According to the sign of the exchange coupling $J$, a parallel ($J>0$) or an antiparallel ($J<0$) alignment of itinerant and localized spin is
favored with remarkable differences in the physical properties. We restrict our considerations in the following to the $J>0$ case, sometimes
referred to as `ferromagnetic Kondo-lattice model (FKLM)' ($s-f$, $s-d$ model) or, in the strong-coupling regime, as the `double-exchange model'. 
The so-called $p-d$ exchange, not considered here, would require $J < 0$ and different, significantly greater, absoluted values\cite{Kacman01}. 
At first glance, one does not expect dramatic changes of the collective magnetic order since in the first order the features follow $J^2$. 
In the following, the charge carriers will be itinerant electrons. The generalization to the case of holes is straightforward. 

The KLM refers to magnetic materials that get their magnetic properties from a system of localized magnetic moments 
being indirectly coupled via interband exchange to itinerant carriers. 
The magnetic properties of the FKLM, therefore, strongly correlate with the electronic properties of the itinerant electron subsystem. 
Thus, the evaluation of the FKLM includes the solution of two interconnected partial problems, 
mainly an electronic problem and a magnetic local-moment problem. 
In addition, for the diluted systems, the inherent positional disorder of the magnetic ions does 
not only break the translational symmetry of the crystal lattices but also influences the electrons distribution. 
The disorder of the magnetic-ion position and the interband coupling greatly complicates the theoretical description of the FKLM, too. 
Therefore, the evalution of the FKLM has to deal with the conduction band disorder and on-site local-moment disorder. 
So it is really difficult to investigate all those factors instantaneously as the same level. 
In this paper, accordingly we apply a strategy of self-consistent calculation -- the procedure is mainly divided into two parts: 
1) the electronic part in which the local-moment magnetization is fixed as parameter; 
2) the local magnetic part in which the solution of the electronic part leads to an effective exchange integral 
   for the interaction between the local spins. 
The following sections \ref{sec:KLM}, \ref{sec:DAA}, \ref{sec:MRKKY} and \ref{sec:EffHM} show the theoretical methods and 
the section \ref{sec:Selfmethod} describes the whole loop of self-consistent calculation. 
In the last analysis, the model description refers to the mechanism of carrier-mediated ferromagnetism 
as it is the case in diluted III-Mn-V and $n$-doped II-Mn-VI samples. 

%%%%%%%%%%%%%%%%%%%%%%%%%%%%%%%%%%%%%%%%%%%%%%%%%%%%%%%%%%%%%%%%%%%%%%%%%%%%%%%%
\subsection{The concentrated Kondo-lattice model}\label{sec:KLM}
%%%%%%%%%%%%%%%%%%%%%%%%%%%%%%%%%%%%%%%%%%%%%%%%%%%%%%%%%%%%%%%%%%%%%%%%%%%%%%%%

The model Hamiltonian (\ref{eq:Hamil}) provokes a rather sophisticated many-body problem. 
Using second quantization, the interaction term (\ref{eq:Hsf}) can be rewritten as: 
\begin{equation}\label{eq:sfsq}
  H_{sf}=-\frac{J}{2}\sum_{j\sigma}(z_{\sigma}S_{j}^{z}n_{j\sigma}+S_{j}^{-\sigma}c_{j\sigma}^{\dagger}c_{j-\sigma}) \;,
\end{equation}
where $z_{\sigma}=\delta_{\sigma\uparrow}-\delta_{\sigma\downarrow}$, $S_{j}^{\sigma}=S_{j}^{x}+{\rm i}z_{\sigma}S_{j}^{y}$ and 
$n_{i\sigma}=c_{i\sigma}^{\dag}c_{i\sigma}$ is the occupation number operator at site $i$. 
The first term on the r.h.s of (\ref{eq:sfsq}) shows an Ising-like interaction between the $z$-components of the spin operators and 
the second term describes spin-exchange processes. 

The electronic part of the many-body problem of the above Hamiltonian is solved as soon as one gets 
the retarded single-particle Green function $G_{{\bf k}\sigma}^{el}(E)$: 
\begin{eqnarray}\label{eq:GF}
G_{ij\sigma}^{el}(E)&\!\!=\!\!&\langle\langle c_{i\sigma};c^{\dag}_{j\sigma} \rangle\rangle_{_E} = 
                \frac{1}{N}\sum_{\bf k}G_{{\bf k}\sigma}^{el}(E) e^{ i{\bf k} \cdot {\bf R}_{ij} } \;, \\
G_{{\bf k}\sigma}^{el}(E)&\!\!=\!\!&\frac{1}{E-\epsilon({\bf k})-M_{{\bf k}\sigma}(E)} \;, 
\end{eqnarray}
where the superscript \textit{``el''} refers to \textit{``electron''}. 
The Bloch energy $\epsilon(\mathbf{k})$ is the Fourier transform of the hopping integral $T_{ij}$:
$\epsilon(\mathbf{k})=N^{-1} \sum_{ij} T_{ij} e^{ -i \mathbf{k}\cdot\mathbf{R}_{ij}}$ and $\mathbf{R}_{ij}={\bf R}_i-{\bf R}_j$. 
The electronic self-energy $M_{{\bf k}\sigma}(E)$ becomes the central quantity of the many-body problem of the electronic part. 

For finite temperatures and arbitrary band occupations, an exact expression of $M_{{\bf k}\sigma}(E)$ is not available and 
one needs to apply an approximation\cite{Nolting96,Nolting97}. The method that we choose is 
the \textit{Interpolating Self-energy Approach} (ISA)\cite{Nolting01}, which results in a wave-vector independent self-energy 
\begin{equation}\label{eq:CSE}
  M_{\sigma}(E) = - \frac{J}{2} z_{\sigma} \langle S^{z} \rangle 
                  + \frac{J^{2}}{4}  \frac{a_{\sigma}  G_{0}^{el}(E-\frac{J}{2}z_{\sigma}\langle S^{z}\rangle)}
                                         {1-b_{\sigma} G_{0}^{el}(E-\frac{J}{2}z_{\sigma}\langle S^{z}\rangle)} .
\end{equation} 
The parameters $a_{\sigma}$, $b_{\sigma}$ are fixed by rigorous high-energy expansions to fulfill the first four spectral moments:
\begin{equation}\label{eq:par}
  a_{\sigma}=S(S+1)-z_{\sigma}\langle S^{z}\rangle(z_{\sigma}\langle S^{z}\rangle+1) \;, \;
  b_{\sigma}=b_{-\sigma}=\frac{J}{2}
\end{equation}
and $G_{0}^{el}(E)$ is the \textit{free} propagator:
\begin{equation}\label{eq:prop}
  G_{0}^{el}(E)=\frac{1}{N}\sum_{ \mathbf{k}}\frac{1}{E-\epsilon(\mathbf{k}) } \;.
\end{equation}
The ansatz (\ref{eq:CSE}) is chosen to be exact for a maximum number of special cases in the low-density limit. 
Therefore, it should represent a reasonable starting point for the description of ferromagnetic semiconductors. 
The ISA approach has already been applied in model studies\cite{Nolting04} and to real substances\cite{Kreissl05}. 
For a detailed account of the decoupling procedure and an extensive discussion of the reliability of this self-energy 
we refer the reader to Ref.\cite{Nolting01}. 
$M_{\sigma}(E)$ is the electronic self-energy for the \textit{concentrated} KLM and, 
in the next subsection, will be used to simulate the disorder of the magnetic moments in diluted ferromagnetic semiconductors. 

%%%%%%%%%%%%%%%%%%%%%%%%%%%%%%%%%%%%%%%%%%%%%%%%%%%%%%%%%%%%%%%%%%%%%%%%%%%%%
\subsection{Dynamical alloy analogy}\label{sec:DAA}
%%%%%%%%%%%%%%%%%%%%%%%%%%%%%%%%%%%%%%%%%%%%%%%%%%%%%%%%%%%%%%%%%%%%%%%%%%%%%

For diluted ferromagnetic semiconductors A$_{c_A}$B$_{c_B}$ of nonmagnetic ions `A' and magnetic ions `B', where $c_A$ and $c_B$ 
refer to the concentrations, one should investigate the effect of disorder on the electronic self-energy. 
If the atomic level of A sites is 
\begin{equation}\label{eq:Aatom}
  \epsilon_{A\sigma}=\varepsilon_{0} \;,
\end{equation}
using dynamical alloy analogy (DAA), the atomic level of B sites reads as a \textit{dynamic} atomic energy level: 
\begin{equation}\label{eq:Batom}
 \epsilon_{B\sigma}=\varepsilon_{0}+M_{\sigma}(E)
\end{equation}
since the local interband exchange $H_{sf}$ acts on the charge carriers. 
Neglecting a Coulomb disorder potential which might be important in some circumstances\cite{Timm03} (e.g. metal-insulator transition), 
the single-particle properties can then be derived from the propagator 
\begin{equation}\label{eq:Rprop}
  R_{\sigma}(E)=\int\limits_{-\infty}^{+\infty} \!\!{\rm d}\omega\, \frac{\rho_{0}(\omega)}{E-\omega-\Sigma_{\sigma}(E)} \;,
\end{equation}
where $\Sigma_{\sigma}(E)$ is now the electronic self-energy in the diluted system and
$\rho_{0}(x)$ the Bloch-density of states of the non-interacting carriers. 
For the determination of the decisive self-energy, one can use a standard CPA formalism \cite{Nolting01}:
\begin{eqnarray}\label{eq:CPA}  
  0 &=& c_A \frac{-\Sigma_{\sigma}(E)}{1-R_{\sigma}(E) (-\Sigma_{\sigma}(E))} \nonumber\\
    &+& c_B \frac{M_{\sigma}(E)-\Sigma_{\sigma}(E)}{1-R_{\sigma}(E) (M_{\sigma}(E) - \Sigma_{\sigma}(E)) } \;,
\end{eqnarray}
where we have chosen $\varepsilon_0 = 0$. 

The  configurational averaging, inherent in CPA, provides translational symmetry 
and therewith site-independent average spin-dependent occupation numbers:
\begin{equation}\label{eq:occupyeq}
  \langle n_{\sigma}\rangle = \int\limits_{-\infty}^{+\infty} \!{\rm d}E\, \frac{ \rho_{\sigma}(E) }{{\rm e}^{\beta(E-\mu)} + 1} \,
                       \equiv \int\limits_{-\infty}^{+\infty} \!{\rm d}E\, f_{-}(E)\rho_{\sigma}(E) \;,
\end{equation}
where $f_{-}(E)$ is the Fermi function, $\mu$ the chemical potential and 
$\rho_{\sigma}(E)$ the quasiparticle density of states of the {\it interacting} particle system:
\begin{equation}\label{eq:QDOS}
  \rho_{\sigma}(E)=-\frac{1}{\pi}{\rm Im}R_{\sigma}(E) \;.
\end{equation}
Thus, for given magnetization $\langle S^z \rangle$ and occupation number of carrier $n = \langle n_\uparrow\rangle + \langle n_\downarrow\rangle$, 
the many-body problem in the diluted system can be solved by the Eqs. (\ref{eq:CSE}), (\ref{eq:Rprop}), (\ref{eq:CPA}) and (\ref{eq:occupyeq}) 
in a self-consistent manner. The single-particle Green function can be expressed as 
\begin{equation}\label{eq:DGF}
G_{ij\sigma}^{el,\Sigma_\sigma}(E)=\frac{1}{N}\sum_{\mathbf{k}} 
                                   \frac{e^{i\mathbf{k}\cdot \mathbf{R}_{ij}}}{E-\epsilon({\mathbf{k}})-\Sigma_{\sigma}(E) } \;.
\end{equation}
The $G_{ij\sigma}^{el,\Sigma_\sigma}(E)$ describes the electronic properties 
of the itinerant electron subsystem in the diluted ferromagnetic semiconductors. 
In the next subsection, we discuss a modified RKKY (MRKKY) theory\cite{Nolting97,Santos02} 
that will connect the magnetic properties and the electronic properties.

%%%%%%%%%%%%%%%%%%%%%%%%%%%%%%%%%%%%%%%%%%%%%%%%%%%%%%%%%%%%%%%%%%%%%%%%%%%%%%%%
\subsection{Modified RKKY}\label{sec:MRKKY}
%%%%%%%%%%%%%%%%%%%%%%%%%%%%%%%%%%%%%%%%%%%%%%%%%%%%%%%%%%%%%%%

The basic idea of the MRKKY\cite{Santos02} is to map the exchange interaction (\ref{eq:Hsf}) 
on an effective spin Hamiltonian of the Heisenberg type: 
\begin{equation}\label{eq:Hsf2Hf}
  H_{sf} \longrightarrow \langle H_{sf}\rangle^{(c)} \equiv H_f =-\sum_{ij} J_{ij}^{eff} \mathbf{S}_i\cdot \mathbf{S}_j
\end{equation}
by averaging out the conduction electron degrees of freedom ($\langle \cdots \rangle^{(c)}$). 
If one writes $H_{sf}$ as: 
\begin{equation}\label{eq:Hsfk}
  H_{sf}=-J\frac{1}{N}\sum_{i\sigma\sigma^{\prime}}\sum_{\mathbf{k}\mathbf{q}}
  e^{-i\mathbf{q}\cdot\mathbf{R}_i}\left(\mathbf{S}_i\cdot
  \boldsymbol{\hat\sigma}\right)_{\sigma\sigma^{\prime}}
  c^{\dagger}_{\mathbf{k}+\mathbf{q}\sigma}c^{}_{\mathbf{k}\sigma^{\prime}}
\end{equation}
with Pauli spin matrices $\boldsymbol{\hat\sigma}$ of the carrier-spin operator, 
the averaging $\langle H_{sf}\rangle^{(c)}$ means to retain operator character in the subspace of the local spins and to calculate the expectation value: 
\begin{equation}\label{eq:averc+cs}
  \bigl\langle c^{\dagger}_{\mathbf{k}+\mathbf{q}\sigma}c^{}_{\mathbf{k}\sigma^{\prime}}\bigr\rangle^{(c)}  
  = \frac{1}{\Xi^{\prime}}\textrm{Tr}\left(e^{-\beta H^{\prime}} c^{\dagger}_{\mathbf{k}+\mathbf{q}\sigma}c^{}_{\mathbf{k}\sigma^{\prime}}\right).
\end{equation}
$H^{\prime}$ has exactly the same structure as the KLM Hamiltonian $H$ Eq.~(\ref{eq:Hamil}), except for the fact that 
the local spin operators are to be considered as c-numbers for the averaging $\langle \cdots \rangle^{(c)}$, 
therefore not affecting the trace. $\Xi^{\prime}$ is the corresponding grand partition function. 
The expectation value Eq.~(\ref{eq:averc+cs}) does not necessarily vanish for $\mathbf{q}\ne 0$ 
and for $\sigma\ne\sigma^{\prime}$, as it would do when averaging in the full Hilbert space of the KLM. 
To calculate Eq.~(\ref{eq:averc+cs}), one introduces a proper \textit{restricted} Green function
\begin{equation}\label{eq:rectGreen}
\hat G^{\sigma^{\prime}\sigma}_{\mathbf{k},\mathbf{k}+\mathbf{q}}(E)
=\bigl\langle\!\bigl\langle c^{}_{\mathbf{k}\sigma^{\prime}};
 c^{\dagger}_{\mathbf{k}+\mathbf{q}\sigma}\bigr\rangle\!\bigr\rangle^{(c)}_{E},
\end{equation}
which has the \textit{normal} definition of a retarded Green function, only the averages have to be done in the Hilbert space of $H^{\prime}$ 
(the details of the averaging procedure being in the Ref.\cite{Santos02}). 
The solution of the equation of motion for the restricted Green function can be iterated up to any desired accuracy, 
however, at the expensive of higher and higher products of local spin operators. 
To retain the operator form (\ref{eq:Hsf2Hf}) one has to truncate the infinite series in a proper way\cite{Nolting97,Santos02}. 
Then, the effective exchange integrals of the Heisenberg-Hamiltonian $H_f$ can be written as:
\begin{equation}\label{eq:effJ}
J_{ij}^{eff}=\frac{J^2}{4\pi}\!\!\sum_{\sigma}\text{Im}\!\!
                         \int_{-\infty}^{+\infty}\!\!\!{\rm d} E f_{-}(E) G^{el,0}_{ij\sigma}(E) G^{el,\Sigma_\sigma}_{ij\sigma}(E) .
\end{equation}
Since $J_{ij}^{eff}$ is a functional of the conduction electrons self-energy $\Sigma_{\sigma}(E)$, 
the effective exchange integrals receive a distinct temperature and band-occupation dependence. 
Neglecting the self-energy $\Sigma_{\sigma}(E)$, $J_{ij}^{eff}$ leads to the conventional RKKY interaction. 
Once getting the exchange integrals $J_{ij}^{eff}$, one can solve the magnetic properties of the KLM 
from the effective Heisenberg Hamiltonian (\ref{eq:Hsf2Hf}). 

%%%%%%%%%%%%%%%%%%%%%%%%%%%%%%%%%%%%%%%%%%%%%%%%%%%%%%%%%%%%%%%%%%%%%%%%%%%%%%%%
\subsection{The disordered Heisenberg model}\label{sec:EffHM}
%%%%%%%%%%%%%%%%%%%%%%%%%%%%%%%%%%%%%%%%%%%%%%%%%%%%%%%%%%%%%%%%%%%%%%%%%%%%%%%%

In this section, a CPA-like approach\cite{Tang0601,BEB} is used to study a disordered Heisenberg model with long-range exchange. 
In the following, we simply introduce this method and apply it to the diluted spin systems. 

To solve the Heisenberg Hamiltonian
\begin{equation}\label{eq:HMHamil}
  H_f=-{\sum_{i,j=1}^{N}}J_{ij}^{eff}\, \mathbf{S}_i\cdot \mathbf{S}_j \;,
\end{equation}
one introduces the retarded magnon Green's function
\begin{equation}\label{eq:GF_Df}
  G_{ij}(E)=\GREEN{S_i^+}{S_j^-}{_E^{ret}}\;
\end{equation}
where $S_i^\pm=S_i^x\pm i S_i^y$. For simplicity we suppress the index `eff' of the exchange integrals. 
It should be mentioned here that the index $i$ and $j$ in the Eq. (\ref{eq:GF_Df}) 
refer to all sites of systems and is different from the $H_f$ in the Eq. (\ref{eq:Hsf2Hf}). 

For a binary spin systems $\mbox{A}_{c_A}\mbox{B}_{c_B}$, the Green's functions' equation of motion 
based on the well-known Tyablikov approximation reads \cite{Tang0601} 
\begin{equation}\label{eq:Eq_Motion2}
  \mathbb{G}_{ij} = \mathcal{G}_i ( \delta_{ij} - {\sum_{m}} \mathbb{J}_{im} \mathbb{G}_{mj} ) \; ,
\end{equation}
if one introduces the $2 \times 2$ matrices
\begin{equation}\label{eq:Green Matrix}
  \mathbb{G}_{ij} = \begin{pmatrix}
    G_{ij}^{AA} & G_{ij}^{AB} \\
    G_{ij}^{BA} & G_{ij}^{BB} \
  \end{pmatrix} = \begin{pmatrix}
    x_iG_{ij}x_j & x_iG_{ij}y_j \\
    y_iG_{ij}x_j & y_iG_{ij}y_j \
  \end{pmatrix} \;,
\end{equation}
\begin{equation*}\label{eq:J Matrix}
  \mathbb{J}_{ij} = \begin{pmatrix}
    J_{ij}^{AA} & J_{ij}^{AB} \\
    J_{ij}^{BA} & J_{ij}^{BB} \
  \end{pmatrix} \;
\end{equation*}
and
\begin{equation*}\label{eq:g Matrix}
  \mathcal{G}_i = \begin{pmatrix}
    x_i g_i^A & 0 \\
    0 & y_i g_i^B \
  \end{pmatrix} = \begin{pmatrix}
    x_i \frac{\sigma_A}{\omega - \omega_A} & 0 \\
    0 & y_i \frac{\sigma_B}{\omega - \omega_B} \
  \end{pmatrix} \;
\end{equation*}
where $x_i+y_i=1$ ($x_i=1$ or $0$ if $i\in$ A or B, respectively), $J^{XY}_{ij}$ the exchange integrals between X sites and Y sites (X,Y $\in$ A or B), 
$\sigma_{\lambda} = \EW{S_{\lambda}^z} / \sigma_0$ ($\lambda$ = A or B, $\sigma_0 = c_A \EW{S_A^z} + c_B \EW{S_B^z}$), $\omega = E / 2 \sigma_0$ and
the `local potential' $\omega_A$ or $\omega_B$ 
\begin{subequations}\label{eq:Virtual Potential}
\begin{eqnarray}
  \omega_A =  c_A \sigma_A ({\sum_{m}} J_{im}^{AA}) + c_B \sigma_B ({\sum_{m}} J_{im}^{AB}) \;, \\
  \omega_B =  c_A \sigma_A ({\sum_{m}} J_{im}^{BA}) + c_B \sigma_B ({\sum_{m}} J_{im}^{BB}) \;.
\end{eqnarray}
\end{subequations}
Here, the influence of neighbor spins by exchange interaction with site $i$ is taken as an \textit{`effective static field'}. 
The random A-type or B-type occupation of any site determines the `local potential' $\omega_A$ or $\omega_B$, respectively.

Thus the matrix propagator $\mathbb{G}_{ij}$ satisfies an equation for a problem with only `local disorder', 
the matrix $\mathbb{J}_{ij}$ being independent of the random numbers. 
Let us now extend the theory to the special case of a diluted magnetic systems we are exclusively interested in here. 
For this purpose, we assume that A is a nonmagnetic ion, i.e., 
the exchange integrals $J_{ij}^{AA}=J_{ij}^{AB}=J_{ij}^{BA}=0$, 
the averaging $\langle S_A^z \rangle=0$ and $\langle S_B^z \rangle = \langle S^z \rangle$. 
In addition, $J_{ij}^{BB}$ describes the exchange integral of magnetic ion B and therefore $J_{ij}^{BB}=J_{ij}^{eff}$ that can be calculated 
by Eq. (\ref{eq:effJ}). Then, one gets $g_i^A=0$, $G^{AA}=0$ and 
\begin{eqnarray}
G_{ij}^{eff}   &=&   g_i^{eff} ( \delta_{ij} - \sum_m J_{im}^{eff} G_{mj}^{eff} ) \;, \\
g_{i}^{eff} &\equiv& y_i g_i^{B} = y_i ( \omega^{\prime} - c_B \sum_m J_{im}^{eff} ) ^{-1} \; .
\end{eqnarray}
Here $G_{ij}^{eff}\equiv G_{ij}^{BB}$ and $\omega^{\prime}=E/(2\langle S^z\rangle)$. 
Thus, the complete averages lead to $\langle \mathbb{G} \rangle = \langle G^{eff} \rangle$. 

Following the locator approach in Ref.\cite{Tang0601}, one introduces the renormalized locator: 
\begin{equation}\label{eq:gammadef}
\gamma=\langle g_i^{eff} / ( 1 - g_i^{eff} U_0 ) \rangle \;,
\end{equation}
where $U_0$ is the diagonal element of the renormalized interactor, which describes the interaction in the effective medium \cite{Tang0601}. 
Thus the CPA self-consistent condition is expressed as \cite{Gonis77}
\begin{equation}\label{eq:Self Consistent}
  \mathbf{\gamma} \equiv \langle G_{ii}^{eff} \rangle = N^{-1} \sum_{\mathbf{k}} G_{\mathbf{k}}^{eff} \;,
\end{equation}
where $G^{eff}_{\mathbf{k}} = [ \gamma^{-1} + U_0 + J^{eff}({\mathbf{k}}) ]^{-1}$ and 
$J^{eff}(\mathbf{k})$ is the Fourier transform of the exchange interaction $J_{ij}^{eff}$. The renormalized locator can be written as
\begin{equation}\label{eq:gamma}
  \gamma(\omega^{\prime})=\frac{c_B}{\omega^{\prime} - c_B \sum_m J_{im}^{eff} - U_0} \;.
\end{equation}

Once one has found the renormalized locator $\gamma$ from self-consistent CPA equations Eq. (\ref{eq:Self Consistent}) and Eq. (\ref{eq:gamma}), 
the magnetization can be expressed by the Callen equation\cite{Callen63}: 
\begin{equation}\label{eq:Callen-form}
\EW{S^{z}}=\frac{(S-\Phi)(1+\Phi)^{2S+1}+(1+S+\Phi)\Phi^{2S+1} }{(1+\Phi)^{2S+1}-\Phi^{2S+1}}
\end{equation}
where the average magnon number can be calculated by
\begin{equation}\label{eq:eqfi}
  \Phi=\int_{-\infty}^{+\infty} d\omega^{\prime} \frac{D(\omega^{\prime})}{e^{2 \langle S^z \rangle\omega^{\prime}/k_B T}-1 } \;.
\end{equation}
where $D(\omega^{\prime}) = - \text{Im}\gamma(\omega^{\prime})/(\pi c_B)$ is the magnon spectral function. 

Now, for a given temperature and good starting value of $\EW{S^{z}}$, one can get the renormalized locator $\gamma$ 
and the magnon spectral function $D(\omega^{\prime})$ from the CPA self-consistent solution.
Furthermore, by using the Callen equation \eqref{eq:Callen-form}, one can calculate new values of magnetization $\EW{S^{z}}$, 
which are re-inserted in $\gamma$.

%%%%%%%%%%%%%%%%%%%%%%%%%%%%%%%%%%%%%%%%%%%%%%%%%%%%%%%%%%%%%%%%%%%%%%%%%%%%%%%%
\subsection{The full self-consistent loop}\label{sec:Selfmethod}
%%%%%%%%%%%%%%%%%%%%%%%%%%%%%%%%%%%%%%%%%%%%%%%%%%%%%%%%%%%%%%%%%%%%%%%%%%%%%%%%

In this subsection, we describe the full self-consistent loop for treating the FKLM in order to incorporate the theoretical methods 
mentioned in the above subsections, where a conducting electronic problem and a local magnetic problem are solved separately. 

\begin{figure}[bt]
\centerline{\includegraphics[width=0.9\linewidth]{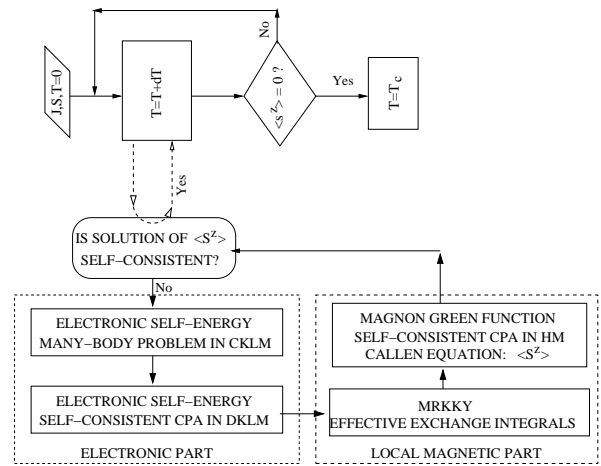}}
\caption{\label{fig:FlowChart}
   Flow chart describing the self-consistent calculation of the diluted ferromagnetic Kondo-lattice model. 
}
\end{figure}
Such a procedure is illustrated in Fig. \ref{fig:FlowChart} and proceeds as follows. 
For a given temperature (e.g. $T=0$) and an initial magnetization $\langle S^z \rangle$, the electronic part of 
the model Hamiltonian (\ref{eq:Hamil}) is solved by using a retarded single-particle Green function within ISA method (in the Sec. \ref{sec:KLM}). 
The result provides the approximate self-energy of the conducting electrons in the concentrated KLM (CKLM). 
In the diluted ferromagnetic systems, for a given occupation of the conduction band, the self-energy of the conduction electrons can be 
determined self-consistently using the coherent potential approximation (CPA) within a \textit{dynamical alloy analogy} (in the Sec. \ref{sec:DAA}). 
To this step, the treatments of the electronic part in the diluted KLM (DKLM) is finished and one gets the single-electron Green function 
of the diluted ferromagnetic systems. In order to study the magnetic properties, we use a modified RKKY (MRKKY) theory that 
results from mapping of the $s-f$ interaction (\ref{eq:Hsf}) onto an effective Heisenberg model (HM) just as it is done in 
the conventional RKKY theory (in the Sec. \ref{sec:MRKKY}). 
The MRKKY theory\cite{Santos02} takes into account higher order terms of the induced conduction electron spin polarization and 
results in an effective exchange integrals $J_{ij}^{eff}$ mainly determined by the electronic self-energy. 
The latter incorporates already the dilution and disorder of the local magnets. 
After using MRKKY theory, we get the effective Heisenberg-exchange integrals between any two lattice-sites. 
Finally, considering random occupation of the magnetic ions, the effective Heisenberg model is solved 
in a CPA framework self-consistently and the new magnetization $\langle S^z \rangle$ can be derived from 
the retarded magnon Green function and the well-known Callen equation (in Sec. \ref{sec:EffHM}). 
The entire procedure of solution is repeated until the solution of $\langle S^z \rangle$ is self-consistent. 
Furthermore, one can change the temperature and repeat the entire $\langle S^z \rangle$ self-consistent calculation until one gets $T_C$. 

%%%%%%%%%%%%%%%%%%%%%%%%%%%%%%%%%%%%%%%%
\section{Magnetic and Electronic Properties}\label{sec:Numerical_Studies}

We have evaluated our theory for a local-moment system ($S=5/2$) on a sc lattice with the lattice constant $a=1$ 
where the width of the free Bloch-band is chosen to be $W=1$ eV. 
In addition, the concentrations of the nonmagnetic and magnetic ions is set to $c_A=1-x$ and $c_B=x$. 
We are interested in how dilution and disorder of the localized moments influence 
the (interconnected) electronic and magnetic properties of the system. 
We start with the inspection of the electronic part in terms of the quasiparticle density of states (QDOS). 

\begin{figure}[htbp]
\centerline{\includegraphics[width=0.9\linewidth]{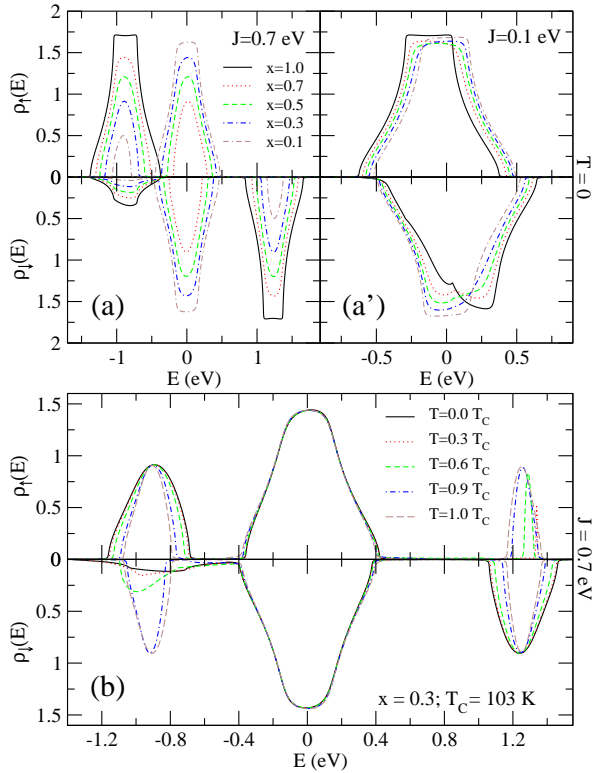}}
\caption{\label{fig:edos}
       Quasiparticle density of states $\rho_\sigma$ of the conduction band as a function of energy in the `diluted' Kondo-lattice model, 
       in (a) and (a') for ferromagnetic saturation ($T=0$) and two different exchange couplings $J$, 
       in (b) for $J=0.7$ eV and different temperatures. 
       Parameters: sc-lattice, $W=1$ eV and $S=5/2$. }
\end{figure} 
Fig.~\ref{fig:edos} (a) and (a') show the quasiparticle density of states of the conduction band at $T=0$ 
for two different exchange couplings $J$ and for several concentrations $x$ of the magnetic ions. 
For not too small exchange couplings $J$ the spectrum is split into several quasiparticle subbands each of them with clear physical meaning. 
The low-energy as well as the high-energy bands in Fig.~\ref{fig:edos} (a) are due to the sites occupied by magnetic moments. 
We denote these bands in the following as \textit{`correlated'} bands. 
The middle structure is due to the non-magnetic sites (\textit{`uncorrelated'} bands). 
Consequently the spectral weight of the uncorrelated band increases with decreasing $x$ at the expense of the weight of the correlated bands. 

Let us first discuss the \textit{`concentrated'} case, i. e. $x=1$. In the saturated ferromagnetic phase ($T=0$),
a spin-up ($\uparrow$) electron has no chance to exchange its spin with the localized moments 
since all spins are parallel to each other and parallel to the electron spin aligned. 
The spin-flip terms in the interaction part of Eq. (\ref{eq:sfsq}) do not work, 
only the Ising-like part (first part of the Hamiltonian $H_{sf}$) 
takes care for a rigid shift by $-\frac{1}{2}JS$ of the $\uparrow$ QDOS compared to the `free' Bloch DOS.

The spin-down $\downarrow$ spectrum is more complex since a $\downarrow$ electron has even at $T=0$ 
two possibilities to exchange its spin with the ferromagnetically saturated local moment system. 
It can emit a magnon therewith reversing its spin. This leads to \textit{`scattering states'} 
which built up the low-energy subband ($\sim -\frac{1}{2}JS$). Because of the spin reversal, 
after which the charge carrier hops over lattice sites with exclusively parallel aligned local-moment spins, 
this subband occupies exactly the same energy region as the $\uparrow$ subband. 

The $\downarrow$ electron has also the possibility to exchange its spin by polarizing its local spin neighborhood 
what can be understood as a repeated magnon emission and reabsorption. 
This creates a typical quasiparticle, which is called \textit{`magnetic polaron'} 
and can be considered as a propagating electron dressed by a virtual magnon cloud. 
In the limit of a single electron in an otherwise empty band the magnetic polaron even gets an infinite lifetime\cite{Nolting96}. 
The high-energy $\downarrow$ subband ($\sim +\frac{1}{2}J(S+1)$) is built up by such polaron states. 
Magnon emission by the $\downarrow$ electron is equivalent to magnon absorption by a $\uparrow$ electron, 
however, only if there are magnons in the system. In ferromagnetic saturation magnons do not exist. 
This is the reason why there is no high-energy part of the $\uparrow$ spectrum at $T=0$. It appears for finite temperatures Fig.~\ref{fig:edos} (b). 

The physical interpretation of the correlated quasiparticle subbands does not change for the \textit{diluted} case $x<1$. 
For not too weak couplings $J$ as $J=0.7$ eV in Fig.~\ref{fig:edos} (a) 
one recognizes the appearance of an additional middle structure ($\sim 0$) the spectral weight of which scales with $1-x$. 
There is only a slight induced exchange splitting of the uncorrelated bands due to hybridization with the correlated subbands. 
For weak couplings as $J=0.1$ eV in Fig.~\ref{fig:edos}(a') correlated and uncorrelated bands 
are mixed preventing a clear interpretation of the various influences. 
With decreasing $x$, the $\uparrow$ and $\downarrow$ spectrum become more and more symmetric 
since the uncorrelated bands are almost unpolarized and the correlated part is merged for low $x$ into the dominating uncorrelated bands. 
This phenomenon shows that for weak $J$ and a given band occupation the dilution will decrease the charge carrier polarization.

Fig.~\ref{fig:edos} (b) shows the temperature-dependent QDOS of the conduction band. 
At finite temperature, there are magnons to be absorbed by the $\uparrow$ electrons and therefore scattering 
and polaron states also appear in the high-energy $\uparrow$ spectrum. 
Consequently, the high-energy $\downarrow$ electron (magnetic polaron) has now a finite spin-flip probability to become a $\uparrow$ electron. 
This spin-flip is not possible for the saturated ferromagnetic phase since there is no high-energy `scattering' part of the $\uparrow$ spectrum. 
In the low-energy subband the electron hops mainly over lattice sites, where it orients its spin parallel to the local moment, 
either without or with preceding spin-flip by magnon emission (absorption). With increasing temperature spin disorder of local moments grows up. 
The average distance for the propagating $\uparrow$ electron to find a parallel local spin therefore becomes larger 
which results in a shrinking width of the lower $\uparrow$ subband with increasing temperature. 
Finally, the spectrum of $\uparrow$ and $\downarrow$ electrons become symmetric at $T\geq T_C$, 
where spin bands are equally populated and spin-polarization disappears. 
The uncorrelated bands do hardly show up any temperature dependence, at least as long as they are separated from the correlated bands.

\begin{figure}[b]
\centerline{\includegraphics[width=0.9\linewidth]{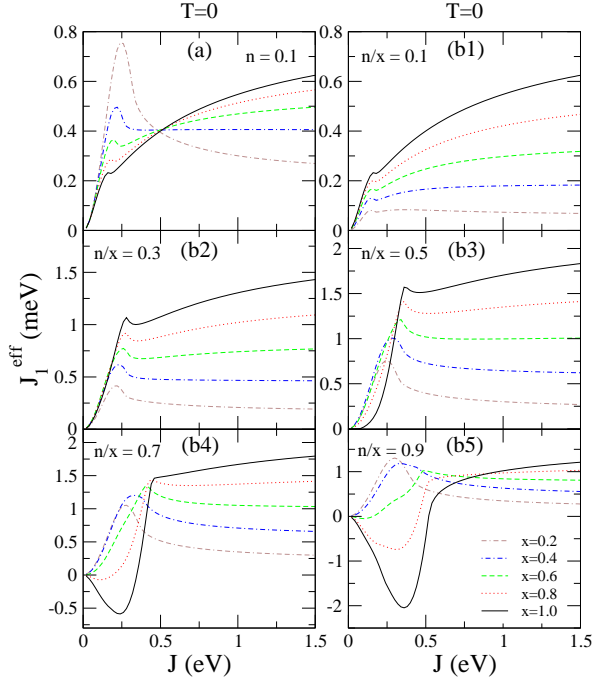}}
\caption{\label{fig:effj_j}
       Effective nearest-neighbor exchange integrals $J_1^{eff}$ at $T=0$ as a function of the interband exchange coupling $J$ 
       and for various moment concentrations $x$: (a) carrier concentration $n=0.1$; 
       (b1)-(b5): average number of electron per magnetic ion: $n/x=$ 0.1,0.3,0.5,0.7,0.9. 
       Parameters: sc-lattice, $W=1$ eV, $S=5/2$. Note the different energy scales. }
\end{figure}
Note that the KLM does not consider a direct exchange between the localized moments. The coupling of the local spins is 
therefore of indirect nature being mainly mediated by the itinerant charge carriers via interband exchange to the localized moments. 
So, the possibility of a collective moment ordering is strongly influenced by the itinerant electron concentration $n$ 
and by the concentration $x$ of the randomly distributed local moments. 
In case of more or less separated correlated and uncorrelated subbands the effective filling of the correlated bands 
is given by $n/x$ which turns out to be the decisive parameter.
Fig.~\ref{fig:effj_j} shows the influence of $x$, $n$ together with the interband exchange coupling $J$ on 
the effective nearest-neighbor Heisenberg-exchange integral $J_1^{eff}$ at $T=0$. The carrier-induced exchange interaction is long-ranged. 
However, by far the most important contribution stems from the nearest-neighbor Heisenberg-exchange integral $J_1^{eff}$. 
It is obvious that the interplay between $J$, $n$ and $x$ does lead to a rather complicated behavior 
of the effective RKKY exchange integrals far beyond the expectation based on conventional RKKY.

In Fig.~\ref{fig:effj_j} (a), we investigate the $J$ dependence of the effective Heisenberg exchange for a given carrier density ($n=0.1$).
For small interband exchange coupling $J\rightarrow 0$ the itinerant charge carriers move almost freely in the environment of local moments. 
Consequently, $J_1^{eff}$ shows a conventional RKKY behavior $J_1^{eff}\propto J^2$ as can be found by perturbation theory. 
For small $J$ and strong dilution (small $x$) the correlated band states are deeply within the uncorrelated band, very close to the band centre. 
Therefore, for small carrier concentrations only uncorrelated states are occupied giving rise to \textit{`conventional coupling behavior'}. 
The $J$ region of \textit{`conventional coupling behavior'} is the larger the stronger the moment dilution, 
because for lower $x$ the system is more `free-electron-like', i.e. less polarized (see Fig.~\ref{fig:edos} (a)). 
This explains why the observed $J_1^{eff}$ maximum increases with decreasing $x$. 
For increasing $J$, the coupling $H_{sf}$ eventually starts to influence decisively the transport of itinerant carriers, e.g. 
by polarizing the spin of the conduction electron. Conventional RKKY is no longer valid.

With $J\rightarrow \infty$, the effective exchange integrals $J_1^{eff}$ saturate. 
Strong coupling $J$ leads to a large gap between the separated correlated and uncorrelated bands (see Fig.~\ref{fig:edos}) 
and a further increase of $J$ will not change remarkably the physics of the KLM. 
For small enough particle concentrations the electron hopping between two magnetic sites now happens no longer 
via the uncorrelated bands as in the weak-coupling case, but almost exclusively via the correlated band. 
On the other hand, the latter are getting narrower with increasing dilution, i.e. decreasing $x$. 
The electron mobility decreases and therewith the carrier-induced coupling between the localized moments.
This may explain why the saturation value of $J_1^{eff}$ is getting smaller with increasing dilution.

In the intermediate $J$ region the problem becomes rather sophisticated because of the competition between 
the different coupling mechanisms working for, respectively, weak and strong interband exchange $J$.

There is no difference Figs.~\ref{fig:effj_j} (a) for $n=0.1$ and Figs.~\ref{fig:effj_j} (b1) 
for fixed $n/x=0.1$ for the concentrated case ($x=1$). 
However, dilution ($x<1$) in Figs.~\ref{fig:effj_j} (b1) means that 
there are less itinerant electrons to provide the coupling between the magnetic moments. 
Different from the situation in Figs.~\ref{fig:effj_j} (a), explained above, the effective nearest-neighbor coupling 
$J_1^{eff}$ decreases in the weak coupling region ($J$ small, conventional RKKY) with increasing dilution. 
The strong-coupling behavior is to be understood as that in Fig.~\ref{fig:effj_j} (a). 

Figs.~\ref{fig:effj_j} (b2)-(b5) demonstrate the change of $J_1^{eff}$ with increasing $n/x$. 
In the strong coupling region, $n/x$ is just the effective occupation of the correlated low-energy subband. 
The disordered KLM is in its double exchange limit favoring parallel spin alignment because that increases the electron hopping probability. 
Thus, the system always has ferromagnetic $J_1^{eff}$ with the highest saturated value at $n/x \sim 0.5$. 
The correlated $\uparrow$ subband is just half-filled, a very convenient situation for ferromagnetism in the KLM\cite{Santos02}. 
The decrease of the strength of $J_1^{eff}$ with decreasing $x$ for fixed $n/x$ is again 
due to the accompanying decrease of the free charge carrier concentration. 
The maximum at $n/x \sim 0.5$ in the strong coupling $J_1^{eff}$-curves is the result of two competing mechanisms. 
At first, increasing $n/x$ for fixed $x$ means that the system gets more itinerant charge carriers, 
so that the strength of $J_1^{eff}$ tends to increase. On the other hand, increasing $n/x$ means also that 
the correlated subband fills up, so that the quasiparticle hopping is going to be blocked due to Pauli principle. 
Accordingly, the tendency to antiferromagnetic moment ordering and respective antiferromagnetic correlations come into play suppressing $J_1^{eff}$. 

The competition between the two counteracting tendencies become even more evident in the intermediate coupling range ($J=0.25\cdots0.75$ eV). 
The correlated subbands are already split off especially for high concentration $x$. 
They are effectively half-filled, i.e. the low-energy part is occupied, the high-energy part is empty, when $n/x$ approaches unity. 
We know from the `normal' KLM that the antiferromagnetic phase as well as 
phase separation become likely\cite{Kienert06,Hennig_T} due to the Pauli principle. 
This manifests itself in negative exchange couplings between the localized moments (Figs.~\ref{fig:effj_j} (b4),(b5)). 
However, antiferromagnetism is very sensitive to local-moment dilution\cite{Mucke_T}. 
In the diluted case ($x<1$), due to the hybrid of the correlated band and the uncorrelated band, the itinerant carriers can 
enter the uncorrelated band although $n/x = 1$, i.e. they are no longer forced to hop exclusively over magnetic sites. 
That weakens the Pauli blocking and enhances mobility of the spin polarized charge carriers. 
That is the reason why $J_1^{eff}$ changes from negative to positive values when $x$ decreases from $1.0$ to $0.6$. 

\begin{figure}[tb]
\centerline{\includegraphics[width=0.9\linewidth]{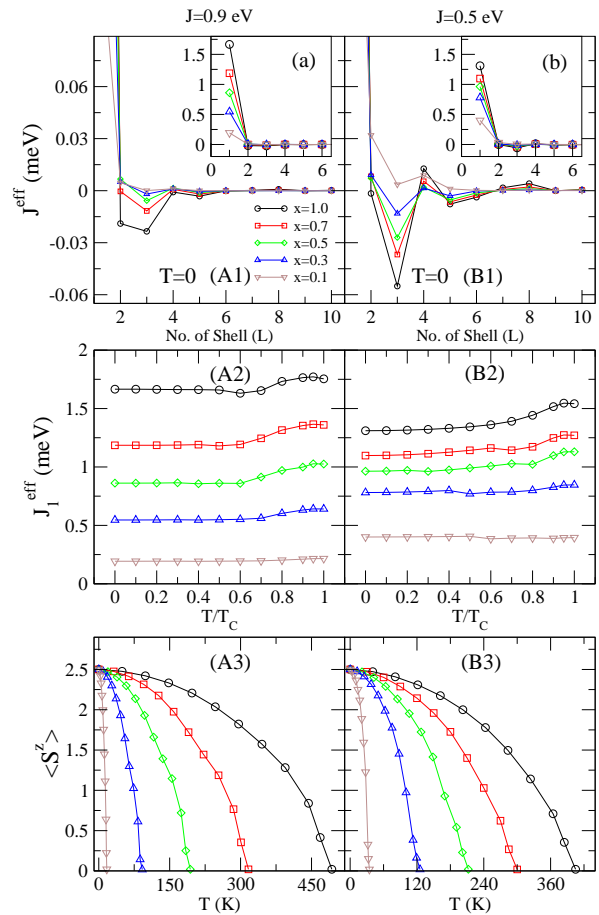}}
\caption{\label{fig:effj_d}
       (A1), (B1): Effective exchange integral $J^{eff}$ at $T=0$ as a function of No. of shell (L) for various concentrations $x$. 
       The L-th shell is built up by L-th next-nearest neighbors. 
       (A2), (B2): Temperature-dependence of the effective nearest-neighbor exchange integral $J_1^{eff}$ for various moment concentration $x$ 
       and two different values of $J$ (0.9 eV and 0.5 eV). 
       (A3), (B3): Magnetization as a function of temperature $T$ for the same concentration $x$ and the same $J$ in the other parts. 
       Parameters: sc-lattice, $W=1$ eV, $S=5/2$, $n=n(T=T_C^{max})$.}
\end{figure}
\begin{figure}[b]
\centerline{\includegraphics[width=0.9\linewidth]{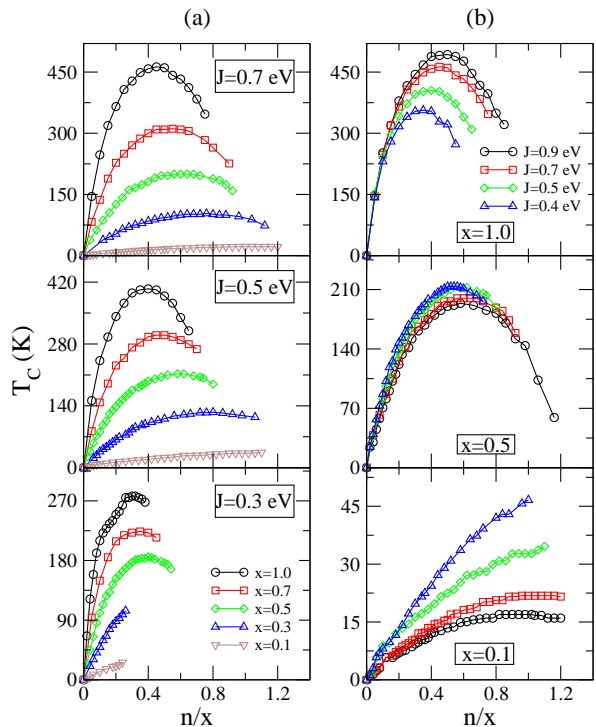}}
\caption{\label{fig:tc_nx}
       Curie temperature $T_C$ as a function of the ratio of the band occupation $n$ and the magnetic ions concentration $x$
       for various interband exchange couplings $J$ and magnetic ions concentration $x$.
       Parameters: sc-lattice, $W=1$ eV, $S=5/2$. }
\end{figure}

Figs.~\ref{fig:effj_d} (A1),(B1) show the distance dependence of the effective exchange integral $J^{eff}$.
The magnetic neighbors of a given magnetic ion are considered as ordered in \textit{`shells'}.
The larger the shell number L the larger the distance to the given magnetic ion. The L-th shell is built up by L-th next-nearest neighbors. 
Since the effective exchange integrals are strongly carrier-concentration dependent the exchange integrals are calculated 
for such $n$-value for which the system has the maximum $T_C$, i.e. $n=n(T=T_C^{max})$ (see Fig.~\ref{fig:tc_nx}). 
The inserts (a) and (b) in Fig.~\ref{fig:effj_d} demonstrate that the effective nearest-neighbor 
$J_1^{eff}$ is much larger than $J_{\rm L}^{eff} ({\rm L}\geq 2)$.
Comparing part (A1) and (B1) of Fig.~\ref{fig:effj_d} one recognizes that $J^{eff}$ originating from 
the weak-coupling interband-exchange (small $J$) is oscillating and long-range corresponding to the well known conventional RKKY-behavior.
For strong $J$ the effective exchange $J^{eff}$ turns out to be substantially stronger damped. 
The magnetism of the FKLM is then sufficiently well described by an effective Heisenberg model with short or intermediate range 
exchange integrals $J^{eff}$ taking into account the first several shells. 
For decreasing $x$, i.e. increasing dilution, $J^{eff}$ has less oscillating behavior. 
For strong dilution $x=0.1$, as realistic for real diluted ferromagnetic semiconductors ($Ga_{1-x}Mn_xAs$), 
all interactions are essentially ferromagnetic over all distances and are also exponentially damped with respect to distance. 
This agrees with recent ab-initio calculations \cite{Bergqvist0405}.

Fig.~\ref{fig:effj_d} (A2), (B2) show the temperature dependence of the effective nearest-neighbor exchange coupling $J_1^{eff}$. 
The $T$ dependence is not very striking except for a slight enhancement near $T=T_C$ that leads to more or less linear behaviors 
of the magnetization $\langle S^z\rangle$ for $T \rightarrow T_C$ (Fig.~\ref{fig:effj_d} (A3), (B3)). 
However, we do not find the remarkably anomalous magnetization behavior as function of temperature as 
predicted by Sarma \textit{et. al.} \cite{SarmaDMFT} and also by reference\cite{Hilbert04}. 
Furthermore, dilution suppresses $J_1^{eff}$ for all temperatures, and its influence obviously increases with increasing interband exchange $J$. 

Fig.~\ref{fig:tc_nx} shows the Curie temperature $T_C$ as a function of the average number of electrons per magnetic ion $n/x$ 
for various exchange couplings $J$ and different magnetic ion concentration $x$. 
In Fig.~\ref{fig:tc_nx} (a), some general features are worth to be mentioned: 
(1) Extremely low carrier concentrations are already sufficient to induce ferromagnetism.
(2) The transition temperature exhibits a strikingly nonmonotonic behavior with a distinct maximum. 
    This points to the importance of compensation effects in real systems for getting optimal Curie temperatures. 
    The number of free carriers should be smaller than the number of magnetic ions at least for not too low $x$. 
(3) There is always a break-down of the $T_C$ curves at a critical $n/x$. For higher values of $n/x$ the system gets negative magnon energies 
    indicating an instability of the ferromagnetic state against antiferromagnetism or spin glass behavior, especially for the weak $J$. 
(4) For a given $n/x$ and a given $J$ the Curie temperature decreases with increasing dilution, 
    whilst the ferromagnetic $n/x$-region increases as is also found by K. Aryanpour et. al. \cite{Aryanpour05}. 
Note that the effects due to dilution and disorder, respectively, are strongly interconnected and are not separated in this paper. 
In our previous paper\cite{Tang0602}, we have demonstrated with a different method 
that dilution appears to be more important for the ferromagnetism than disorder. 

Part (b) of Fig.~\ref{fig:tc_nx} shows $T_C$ as a function of $n/x$ for various $x$ and $J=0.4\cdots0.9$ eV. 
The general features are the same as in the left part. However, it is interesting to observe 
that for the non-diluted case $x=1.0$ larger exchange coupling $J$ leads to higher $T_C$, as intuitively expected. 
For a moment concentration of about $x=0.5$ the $T_C$-curves appear to be almost independent of $J$, 
while for strong dilution $x=0.1$ stronger $J$ lead to lower transition temperatures. 
This is a direct consequence of the nonmonotonic behavior of the effective exchange coupling 
between the randomly distributed localized magnetic moments, discussed above. 
In the highly diluted case $x=0.1$ and for strong couplings $J$ the low-energy correlated band is well separated from 
the broad uncorrelated band being itself rather narrow. 
The `free' carriers are therefore rather immobile since the chemical potential lies within the low-energy quasiparticle subband. 
The carrier immobility leads to a short-range $J^{eff}$ in the strong $J$ case. 
For the strong dilution $x\sim 0.1$, however, the average distance between the local moments is large and 
therefore the short-range $J^{eff}$ has only small influence on $T_C$. 
This is different for smaller $J$ because the correlated band then melts into the uncorrelated one. 
That means a distinctly higher mobility of the charge carriers which mediate the indirect coupling between the local moments 
and therefore leads to a long-range $J^{eff}$. The long-range $J^{eff}$ will effectively enhance $T_C$. 
So we understand that counterintuitively the lower $J$ gives higher $T_C$ than the larger $J$ in the strongly diluted limit. 
The absolute values of $T_C$ are of course strongly reduced with decreasing moment concentration.
The situation is completely different for the concentrated case $x=1.0$ because then no uncorrelated band exists, 
where \textit{`normal'} behavior is observed, namely higher $T_C$ for stronger interband exchange $J$. 
For $x=0.5$ both trends are obviously cancelling each other. 
The DMFT calculation by Chattopadhyay \textit{et. al.} \cite{SarmaDMFT} shows also a nonmonotonic behavior of the transition temperature. 
In their calculation, the weak $J$ broadens the ferromagnetic regions even up to $n/x \sim 3$, probably due to an artefact of the method. 
On the other hand, the ferromagnetic regions for the strong $J$ are limited to $n/x < 1$, rather similar to our results. 
Furthermore, the observed $T_C$-reduction for strong $J$ in the very diluted case applies for $n/x \sim 1$ in their calculation as well. 

\section{Summary}\label{sec:Summary}

We have performed a selfconsistent model calculation of the electronic and magnetic properties 
of diluted local-moment systems described by an extended (ferromagnetic) Kondo-lattice model (s-f model). 
It was the main goal to treat the exchange interaction and the moment disorder on the same level. 
The electronic selfenergy was derived by a Green's function formalism (\textit{`Interpolating Selfenergy Approach'} ISA) 
previously developed and tested for the \textit{`normal'} KLM. 
The ISA had to be generalized to the diluted case, i.e. to the situation of randomly distributed magnetic moments. 
For this purpose we applied a CPA treatment to a dynamical alloy analogy of the diluted KLM. 
The resulting self-energy turns out to be dependent on the magnetization of the diluted and disordered local-moment system. 
This quantity was determined by mapping the interband exchange interaction of the KLM on an effective Heisenberg model. 
The effective Heisenberg-exchange integrals appear as functionals of the electronic selfenergy getting 
therewith a distinct temperature- and carrier concentration dependence. 
In the last step we solved the problem of the diluted Heisenberg spin system by a generalized locator-CPA. 
Finally we arrived at a closed system of equations which could be solved selfconsistently for 
the electronic and magnetic properties of the diluted local-moment system. 
The latter could be discussed in dependence of the moment concentration $x$, 
the concentration of itinerant charge carriers $n$, the interband exchange coupling $J$, and the temperature $T$.

The electronic quasiparticle energy spectrum is strongly influenced by the dilution and disorder of the local-moment system. 
For strong enough exchange $J$ and finite temperature $T$ the quasiparticle density of states consists for each spin direction of three subbands. 
Two of them (\textit{`correlated subbands'}) are due to the existence of magnetic moments and 
are responsible for characteristic correlation effects which manifest themselves, e.g., in a remarkable temperature dependence. 
The third (\textit{`uncorrelated'}) subband, which exhibits hardly any observable $T$-dependence, is caused by the nonmagnetic sites. 
For weak couplings the three subbands are hybridizing. 
Since quasiparticle density of states and electronic selfenergy, respectively, decisively influence the effective Heisenberg exchange integrals, 
the special electronic quasiparticle structure determines the magnetic properties, too. 

The effective Heisenberg exchange integrals reproduce in the weak coupling limit ($J\rightarrow 0$) 
and for not too strong dilutions the well known RKKY-behavior ($J_{ij}\propto J^2$), i.e. long range and oscillating. 
This changes drastically, however, for moderate and strong couplings, where due to electronic correlation effects a strongly 
non-monotonic $J$-dependence does appear. When the average number of electrons per magnetic ion $n/x$ approaches $1$, 
the dominating effective nearest-neighbor exchange integral $J_{ij}^{eff}$ assumes negative values indicating antiferromagnetic tendencies. 
It is worth to be noted that the approach covers RKKY as well as double exchange behavior. 

The Curie temperature $T_C$ of the diluted spin system has been calculated as a function of the average number of electrons 
per magnetic ion $n/x$ for various exchange couplings $J$ and different magnetic ion concentrations $x$. 
We found that extremely low carrier concentrations are already sufficient to induce ferromagnetism, 
where the transition temperature exhibits a strikingly non-monotonic behavior with a distinct maximum. 
This demonstrates the importance of compensation effects in real systems for getting optimal Curie temperatures.
The number of free carriers should be smaller than the number of magnetic ions at least for not too low $x$. 
The results also show that there is always a break-down of the $T_C$ curves at a critical $n/x$, 
indicating instabilities of the ferromagnetic state against antiferromagnetism or spin glass phases. 
For a given $n/x$ and a given $J$ the Curie temperature decreases with increasing dilution, 
although the ferromagnetic $n/x$-region simultaneously broadens. 
As far as the dependence of $T_C$ on $J$ (for a given $n/x$) is concerned we observe two regimes. 
For high $x$ a higher $J$ means an increased Curie temperature. However, for lower $x$ an increasing $J$ suppresses $T_C$. 

Our investigation of the mutual influence of magnetic correlations and disorder effects in diluted local-moment 
systems clearly indicates for all interesting parameter constellations the necessity to treat both 
phenomena on comparable theoretical level, and that beyond conventional perturbation or mean field theory. 
The complexity of the problem did not yet allow us to include the inspection of antiferromagnetic phases 
which is intended for the near future. To tackle real diluted ferromagnetic semiconductors such as $Ga_{1-x}Mn_xAs$ 
it is also planned to combine our model study for realistic parameters with an `ab initio' band structure calculation 
to allow for a direct comparison with experimental data. 

\begin{acknowledgments}
We thank Mr. Jochen Kienert and Mr. Vadym Bryksa for helpful discussions. 
G.X.T is supported by the State Scholarship Programs of China Scholarship Council 
and also benefit from the project (No. 50375040) supported by National Natural Science Foundation of China 
and the project (No. HIT.MD2002.16) supported by Foundation of Harbin Institute of Technology.
\end{acknowledgments}

%%%%%%%%%%%%%%%%%%%%%%%%%%%%%%%%%%%%%%%%%%%%%%%%%%%%%%%%%%%%%%%%%%%%%%%%%%%

%\bibliography{literature}

\end{document}